\documentclass{article}

\usepackage{arxiv}
\PassOptionsToPackage{numbers}{natbib}
\usepackage[utf8]{inputenc} 
\usepackage[T1]{fontenc}    
\usepackage{hyperref}       
\usepackage{url}            
\usepackage{booktabs}       
\usepackage{amsmath,amssymb,amsfonts}
\usepackage{nicefrac}       
\usepackage{microtype}      
\usepackage{lipsum}		
\usepackage{graphicx}
\usepackage{natbib}
\usepackage{doi}
\usepackage{tabularx}
\usepackage{textcomp}
\usepackage{hyperref} 
\usepackage{algorithmic}

\title{Quantum-Enhanced Channel Mixing in RWKV Models for Time Series Forecasting}

\date{} 					

\author{%
  \href{https://orcid.org/0000-0003-0807-0217}{\includegraphics[scale=0.06]{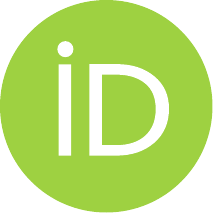}\hspace{1mm}Chi-Sheng Chen\thanks{Corresponding author}} \\
  Neuro Industry Research\\
  Neuro Industry, Inc.\\
  Boston, MA, USA \\
  \texttt{m50816m50816@gmail.com} \\
  \texttt{cchen34@bidmc.harvard.edu} \\
  \And
  \href{https://orcid.org/0000-0002-6770-0285}{\includegraphics[scale=0.06]{orcid.pdf}\hspace{1mm}En-Jui Kuo} \\
  Department of Electrophysics\\
  National Yang Ming Chiao Tung University\\
  Hsinchu, Taiwan \\
  \texttt{kuoenjui@nycu.edu.tw} \\
}

\renewcommand{\shorttitle}{\textit{arXiv} Template}

\hypersetup{
pdftitle={A template for the arxiv style},
pdfsubject={cs.CV},
pdfauthor={Chi-Sheng ~Chen},
pdfkeywords={Deep learning},
}

\begin{document}
\maketitle

\begin{abstract}
Recent advancements in neural sequence modeling have led to architectures such as RWKV, which combine recurrent-style time mixing with feedforward channel mixing to enable efficient long-context processing. In this work, we propose QuantumRWKV, a hybrid quantum-classical extension of the RWKV model, where the standard feedforward network (FFN) is partially replaced by a variational quantum circuit (VQC). The quantum component is designed to enhance nonlinear representational capacity while preserving end-to-end differentiability via the PennyLane framework.

To assess the impact of quantum enhancements, we conduct a comparative evaluation between QuantumRWKV and its classical counterpart across ten synthetic time-series forecasting tasks, encompassing linear (ARMA), chaotic (Logistic Map), oscillatory (Damped Oscillator, Sine Wave), and regime-switching signals. Our results show that QuantumRWKV outperforms the classical model in 6 out of 10 tasks, particularly excelling in sequences with nonlinear or chaotic dynamics, such as Chaotic Logistic, Noisy Damped Oscillator, Sine Wave, Triangle Wave, Sawtooth, and ARMA. However, it underperforms on tasks involving sharp regime shifts (Piecewise Regime) or smoother periodic patterns (Damped Oscillator, Seasonal Trend, Square Wave).

This study provides one of the first systematic comparisons between hybrid quantum-classical and classical recurrent models in temporal domains, highlighting the scenarios where quantum circuits can offer tangible advantages. We conclude with a discussion on architectural trade-offs, such as variance sensitivity in quantum layers, and outline future directions for scaling quantum integration in long-context temporal learning systems.

The code can be obtained on GitHub: \underline{https://github.com/ChiShengChen/QuantumRWKV}.
\end{abstract}


\section{Introduction}
Neural models for time series forecasting have become increasingly important across scientific and industrial applications, including nutrition \cite{chen2025improving}, neuroscience \cite{chen2024quantum}, and dynamical system modeling \cite{chen2025unraveling}. While Transformer-based architectures \cite{vaswani2017attention} have achieved remarkable success in natural language processing, their reliance on self-attention mechanisms incurs quadratic time and memory complexity, limiting their applicability in long-context temporal scenarios. RWKV \cite{peng2023rwkv}, a recent attention-free architecture, addresses this limitation by combining a recurrent-style time-mixing mechanism with feedforward channel mixing, achieving both scalability and competitive performance. However, the expressiveness of its feedforward network (FFN) component remains constrained by the capabilities of classical neural layers.

Motivated by the emerging potential of quantum neural networks (QNNs) \cite{biamonte2017quantum} to represent complex entangled states and nonlinear transformations in high-dimensional Hilbert spaces, we introduce QuantumRWKV, a hybrid quantum-classical extension of the RWKV architecture. In QuantumRWKV, we retain the classical RWKV time-mixing structure while replacing the FFN in the channel mixing module with a dual-branch design: a standard classical multilayer perceptron (MLP) \cite{mcculloch1943logical} branch and a quantum branch consisting of a variational quantum circuit (VQC). The quantum path encodes inputs using angle embedding, applies entangling layers, and returns expectation values of Pauli-Z observables, which are integrated with the classical output and gated via a learned receptance mechanism. This design allows quantum computation to enhance the nonlinear transformation capacity of the model while remaining fully differentiable and compatible with PyTorch \cite{ketkar2021introduction} training pipelines via the PennyLane \cite{bergholm2018pennylane} framework.

To evaluate the effectiveness of this hybrid model, we conduct a comparative study between QuantumRWKV and its purely classical counterpart on a suite of synthetic time-series forecasting tasks. These tasks are designed to probe different dynamic regimes, including linear autoregressive (ARMA) \cite{makridakis1997arma}, chaotic (Logistic Map) \cite{may1987chaos}, oscillatory (Damped and Noisy Oscillator), piecewise-regime (Piecewise and Square Wave), and periodic (Sine, Triangle, and Seasonal Trend) signals. Our experiments show that QuantumRWKV outperforms the classical model in tasks involving smooth nonlinearity, chaotic dynamics, or noise-driven signals, such as Chaotic Logistic, Noisy Damped Oscillator, and Sine Wave. In contrast, classical RWKV performs better in sharply discontinuous tasks like Piecewise Regime and Square Wave, where the discrete nature of transitions may not align well with the smooth encoding of variational quantum circuits.

These findings suggest that quantum augmentation is particularly beneficial for modeling continuous, entangled, or nonlinear dynamics in temporal sequences, while classical structures remain more robust for abrupt or piecewise behaviors. This work provides one of the first controlled, task-specific comparisons between quantum-enhanced and classical recurrent models in the time-series domain, and highlights practical conditions under which quantum integration provides a measurable advantage. Through this study, we aim to clarify the role of quantum circuits in temporal neural modeling and to pave the way for further research on hybrid architectures that combine classical scalability with quantum expressiveness.

The contributions of this work are stated as follows:
\begin{itemize}
    \item \textbf{Hybrid Architecture Design:} We propose \textbf{QuantumRWKV}, a hybrid quantum-classical recurrent model that augments the RWKV architecture by embedding a variational quantum circuit (VQC) into the channel mixing (feedforward) pathway. This design preserves the model’s scalability while enhancing its nonlinear transformation capabilities.
    
    \item \textbf{End-to-End Trainable Quantum Integration:} Our architecture maintains differentiability and compatibility with PyTorch training pipelines using the PennyLane framework, enabling seamless quantum-classical co-training without handcrafted quantum features.
    
    \item \textbf{Controlled Empirical Evaluation:} We perform a task-specific evaluation on ten synthetic time-series forecasting tasks that vary in linearity, periodicity, chaotic behavior, and discontinuities, offering a rigorous benchmark to assess the quantum advantage.
    
    \item \textbf{Task-Specific Insights:} QuantumRWKV consistently outperforms the classical RWKV in tasks characterized by smooth nonlinear dynamics, chaotic behavior, and noise-driven variations (e.g., Chaotic Logistic, Noisy Damped Oscillator, Sine Wave, Triangle Wave, Sawtooth, ARMA). In contrast, classical RWKV exhibits better performance on tasks involving sharp regime transitions or simple deterministic structures (e.g., Piecewise Regime, Damped Oscillator, Seasonal Trend, Square Wave).

    \item \textbf{Theoretical and Practical Implications:} Our results offer insight into how quantum circuits can contribute to temporal modeling, clarifying their limitations in representing sharp regime shifts and guiding future design of quantum-enabled neural architectures.
\end{itemize}

\section{Related work}
\label{sec:headings}


\subsection{Recurrent-Free Sequence Models}
The Transformer architecture has become the de facto standard for sequence modeling in natural language processing (NLP), but its quadratic attention mechanism is inefficient for long time-series tasks. To address this, recent architectures such as Performer \cite{choromanski2020rethinking}, Linformer \cite{wang2020linformer}, and S4 \cite{gu2021efficiently} have proposed alternatives to global attention. RWKV \cite{peng2023rwkv} further simplifies the architecture by completely removing attention, using a recurrent-style time-mixing operation to capture long-range dependencies with linear complexity. In our work, we build on RWKV as the temporal backbone due to its efficiency and scalability in long-context modeling, and explore how quantum-enhanced channel mixing can further improve expressiveness.

\subsection{Quantum Neural Networks and Hybrid Architectures}
Quantum neural networks (QNNs) are a class of models that leverage variational quantum circuits (VQCs) to perform nonlinear transformations in Hilbert space. These models have been shown to be expressive universal approximators \cite{chen2022quantum} and have demonstrated potential advantages in learning complex decision boundaries in low-data regimes \cite{goto2021universal, chen2025quantumgen}. Hybrid quantum-classical networks, where only parts of a model are quantum (e.g., a layer or module), are particularly attractive in the Noisy Intermediate-Scale Quantum (NISQ) era due to hardware constraints. Notable hybrid models include quantum classifiers \cite{biamonte2017quantum, chen2021end}, quantum GANs \cite{situ2020quantum}, and quantum-enhanced convolutional \cite{chen2024qeegnet, chen2025exploring} or transformer layers \cite{chen2025quantum}. Our work contributes to this growing line by embedding a VQC into a temporal neural model—specifically, the channel mixing layer of RWKV.

\subsection{Quantum Machine Learning for Time Series}
Although most QML studies focus on classification or generative tasks, time-series modeling is an emerging area of interest. Quantum recurrent networks \cite{li2023quantum}, quantum reservoir computing \cite{govia2021quantum}, and parameterized quantum circuits \cite{chen2022quantum} have been explored for predicting classical dynamical systems, though many rely on handcrafted quantum features or small-scale toy models. In contrast, we design an end-to-end differentiable hybrid architecture that incorporates a quantum circuit as a direct component of a temporal model, and benchmark its effectiveness on multiple types of synthetic signals. To our knowledge, this is one of the first works to integrate QML into RWKV and to conduct a systematic comparison across distinct temporal regimes.

\begin{figure}
    \centering
    \includegraphics[width=1\linewidth]{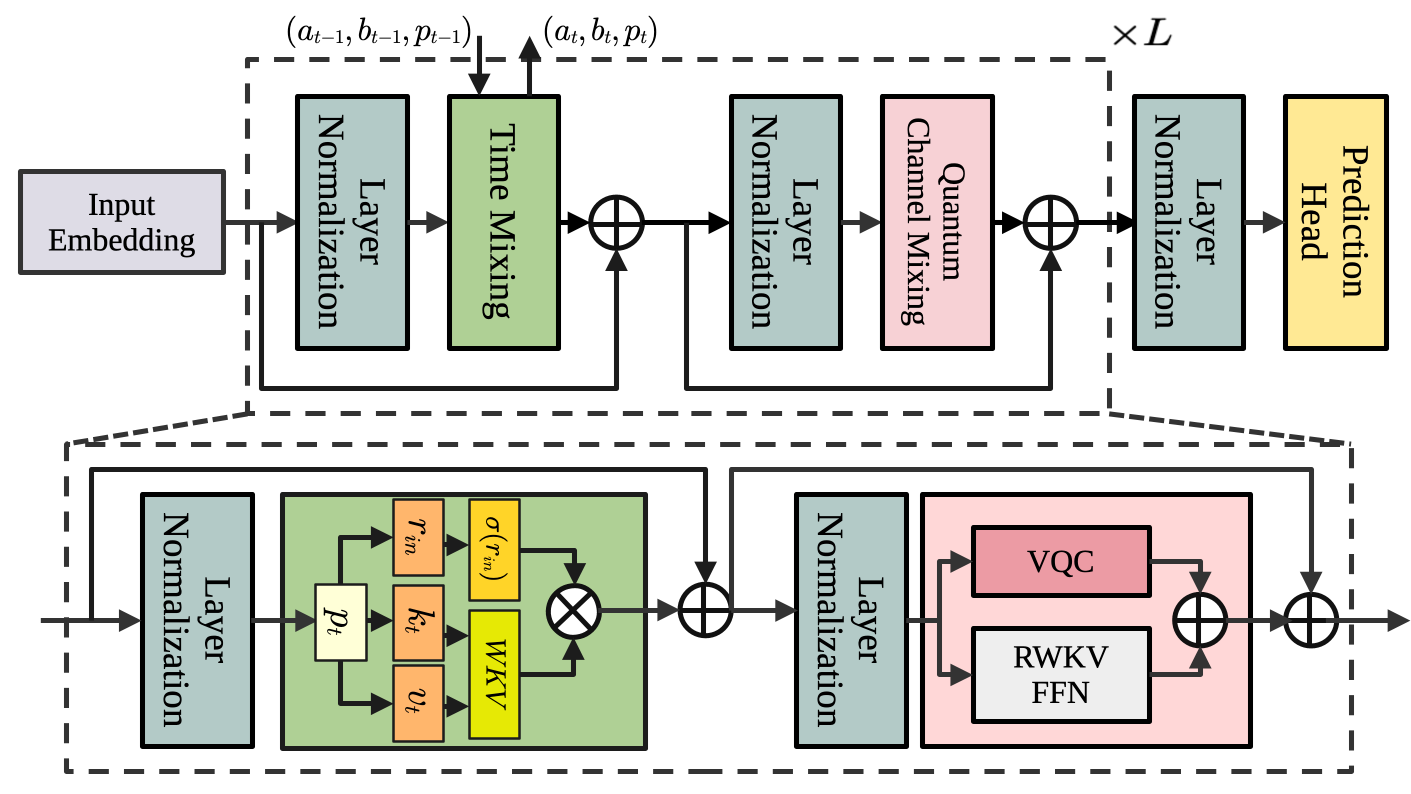}
    \caption{
Architecture of a single QuantumRWKV layer. The model consists of a time mixing block (green) and a quantum-enhanced channel mixing block (pink), each preceded by Layer Normalization and connected via residual paths. The time mixing module performs recurrent weighted key-value (WKV) computations using temporal memory states \((a_t, b_t, p_t)\), which are updated across timesteps and passed between layers. The channel mixing block combines outputs from a classical feedforward path and a variational quantum circuit (VQC), gated by a learned receptance. The final output is normalized and passed to the prediction head.
}
    \label{fig:enter-label}
\end{figure}

\section{Methodology}
\label{sec:others}
In this section, we describe the proposed \textbf{QuantumRWKV} model, a hybrid quantum-classical extension of the RWKV architecture. The model is composed of stacked blocks, each containing a time-mixing module and a channel-mixing module. The key novelty lies in replacing the classical feedforward network (FFN) in the channel-mixing module with a \emph{variational quantum circuit (VQC)}. This hybridization enhances the model’s ability to capture complex nonlinear dynamics in temporal sequences.

\subsection{Classical RWKV Architecture}

The original RWKV model replaces the attention mechanism in Transformers with a time-decay recurrent formulation. At each layer, the model consists of two submodules: a \emph{time-mixing module} (akin to attention but attention-free) and a \emph{channel-mixing module} (equivalent to a position-wise FFN).

\paragraph{Time-Mixing.} 
Given input tensor $x \in \mathbb{R}^{B \times T \times C}$, where $B$ is batch size, $T$ is sequence length, and $C$ is embedding dimension, the time-mixing module computes key $k_t$, value $v_t$, and receptance $r_t$ at each timestep $t$, and uses a recurrence mechanism:
\begin{equation}
\text{WKV}_t = \frac{e^{p_t - \max(p_t, k_t)} \cdot a_{t-1} + e^{k_t - \max(p_t, k_t)} \cdot v_t}{e^{p_t - \max(p_t, k_t)} \cdot b_{t-1} + e^{k_t - \max(p_t, k_t)}}
\end{equation}
Here, $p_t = u + k_t$, with $u \in \mathbb{R}^C$ as a learned time offset, and $a_t, b_t$ as recursive accumulators:
\begin{align}
a_t &= e^{w} \cdot a_{t-1} + v_t \\
b_t &= e^{w} \cdot b_{t-1} + 1
\end{align}
The final output is gated:
\begin{equation}
y_t = \sigma(r_t) \odot \text{WKV}_t
\end{equation}

\paragraph{Channel-Mixing (Classical).} 
In the classical RWKV, channel-mixing is implemented by a gated FFN:
\begin{equation}
h = \sigma(r) \odot W_2 \left( \text{ReLU}(W_1(x))^2 \right)
\end{equation}
where $W_1 \in \mathbb{R}^{C \times d}$, $W_2 \in \mathbb{R}^{d \times C}$, and $d$ is the intermediate FFN dimension.

\subsection{Quantum Channel Mixing}

\begin{figure}
    \centering
    \includegraphics[width=1\linewidth]{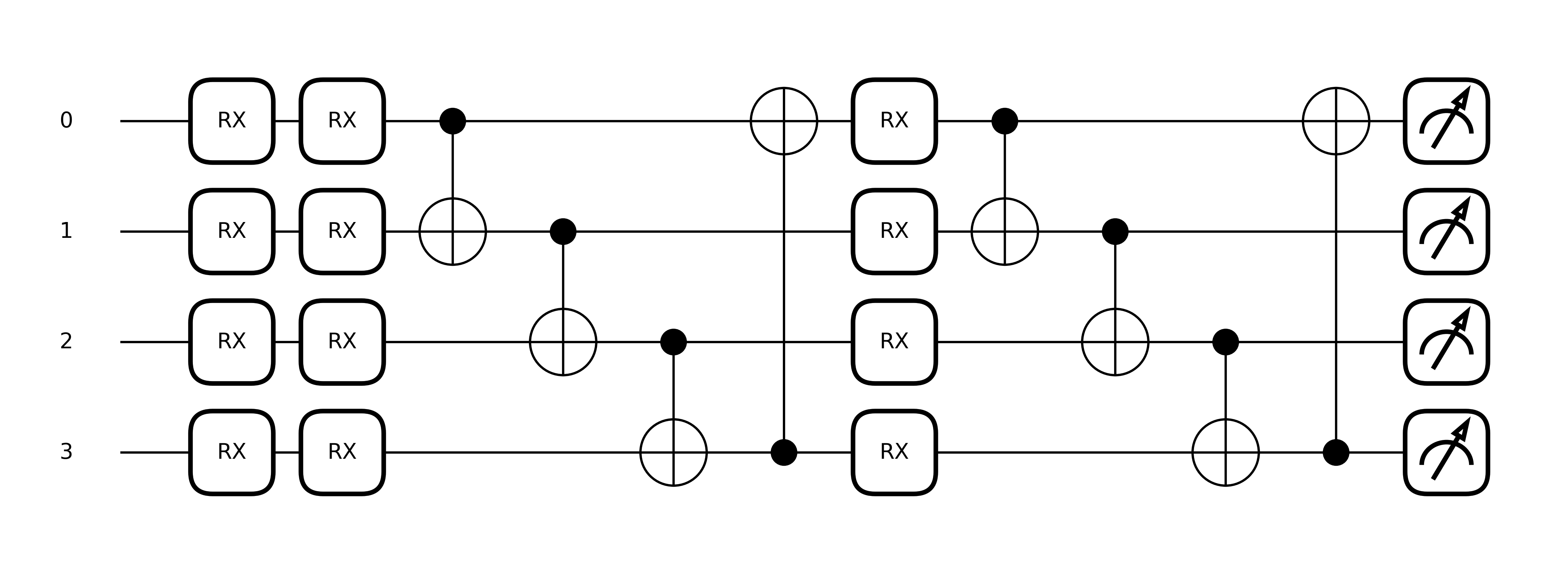}
    \caption{
Structure of the variational quantum circuit (VQC) used in the QuantumChannelMixing module of QuantumRWKV. The circuit operates on 4 qubits and consists of two layers of angle-encoded $R_X(\theta)$ rotations followed by full entangling layers implemented using CNOT gates in a ladder pattern. Each layer encodes the input features into quantum states, and the output is obtained by measuring the expectation values of the Pauli-Z operator on each qubit. Circuit weights are trained jointly with the rest of the model in an end-to-end fashion via backpropagation using the \texttt{default.qubit} simulator in PennyLane.
}

    \label{fig:enter-label}
\end{figure}

To improve non-linear representation power, we replace the classical FFN with a hybrid design that includes a \emph{variational quantum circuit (VQC)}.

\paragraph{Input Projection.}
We project the input to a quantum embedding space:
\begin{equation}
x_q = W_q x, \quad x_q \in \mathbb{R}^{B \times T \times n}
\end{equation}

\paragraph{Variational Quantum Circuit.}

The quantum branch is implemented as a variational quantum circuit (VQC) comprising three main stages: encoding, entanglement, and measurement. First, the input vector is projected to the quantum latent space and encoded via \textit{angle embedding}, where each component $x_{q,i}$ is mapped to the rotation angle of a Pauli-$X$ gate, i.e., $R_X(\theta_i = x_{q,i})$, applied independently to each qubit $i$. This initialization transforms the classical input into a quantum state within the Hilbert space.

The encoded state is then processed through a sequence of entangling layers composed of parameterized quantum gates with trainable weights $\phi \in \mathbb{R}^{L \times n}$, where $L$ is the circuit depth and $n$ is the number of qubits. These layers induce quantum correlations between qubits and serve as the learnable component of the circuit.

Finally, the output is obtained by measuring the expectation values of the Pauli-$Z$ operator on each qubit, yielding a vector $z \in \mathbb{R}^n$ where each element is computed as:
\begin{equation}
z_i = \langle \psi | Z_i | \psi \rangle
\end{equation}
This output vector is subsequently projected back to the model embedding space using a learnable linear transformation, and fused with the classical branch for downstream processing.

The VQC outputs:
\begin{equation}
z = \text{VQC}(x_q; \phi) \in \mathbb{R}^n, \quad z' = W_o z \in \mathbb{R}^C
\end{equation}

\paragraph{Fusion.}
The final output is:
\begin{equation}
\text{QuantumEnhancedChannelMix} = \sigma(r) \odot \left( W_2 \left( \text{ReLU}(W_1(x))^2 \right) + z' \right)
\end{equation}

\subsection{Full Model: QuantumRWKV}

The complete QuantumRWKV stacks $L$ blocks of the form:
\begin{equation}
x_{l+1} = x_l + a_l + \text{QuantumEnhancedChannelMix}(\text{LN}(x_l + a_l)), \quad a_l = \text{TimeMix}(\text{LN}(x_l))
\end{equation}

The model supports both token-based and waveform-based inputs via input projection or embedding layers and concludes with a linear output head.

\section{Experiments}
\label{sec:guidelines}

To evaluate the performance and behavior of the proposed QuantumRWKV model, we conduct experiments on a suite of synthetic time-series forecasting tasks. These tasks are designed to span a diverse range of signal types, including periodic, chaotic, discontinuous, and noisy regimes. The goal is to assess under which conditions the hybrid quantum-classical model exhibits advantages over its classical counterpart.

\subsection{Benchmark Tasks}

To systematically evaluate the effectiveness of QuantumRWKV, we construct a diverse set of synthetic one-dimensional time series tasks designed to capture a range of temporal dynamics. These include both linear and nonlinear, as well as smooth and discontinuous signal types. The benchmark suite comprises: (1) an Autoregressive Moving Average (ARMA) process of order (2,2), modeling linear short-term dependencies; (2) the Chaotic Logistic Map, defined by the recurrence relation $x_{t+1} = r x_t (1 - x_t)$ with $r = 3.9$, which introduces sensitive dependence on initial conditions; (3) a Damped Oscillator, representing a smooth sinusoidal signal with exponential decay; and (4) a Noisy Damped Oscillator, formed by adding Gaussian noise to the previous signal to simulate real-world perturbations. 

To test the model's ability to handle abrupt transitions, we include a Piecewise Regime signal, in which the data-generating rule switches discretely across time. Additionally, we incorporate a set of canonical periodic waveforms—Sine, Triangle, Square, and Sawtooth waves—to assess generalization to structured but varied frequency-domain characteristics. Lastly, we include a Seasonal Trend signal, which combines a linear drift with superimposed periodic fluctuations.

Each sequence consists of 200 timesteps and is partitioned into 80\% training and 20\% testing data. The forecasting objective is to predict the next value $x_{t+1}$ given the history of previous values $x_{1:t}$, under a sliding window setup.

\subsection{Model Configuration}

We use identical architectural settings for both the classical and quantum models, with the only difference being the implementation of the channel-mixing module. The embedding dimension is set to $n_{\text{embd}} = 768$, with a total of $n_{\text{layer}} = 6$ stacked blocks. Each feedforward network (FFN) uses an intermediate projection size of $n_{\text{intermediate}} = 3072$, and the time-mixing mechanism is configured with $n_{\text{head}} = 12$ attention-free heads. For the quantum-enhanced model, the variational quantum circuit employs $n_{\text{qubits}} = 4$ qubits and a circuit depth of $q_{\text{depth}} = 2$. All other architectural components, including layer normalization and output projection, are kept consistent across both model variants.

Quantum circuits are simulated using \texttt{default.qubit} backend. All experiments are run on CPU or single-GPU settings. Each experiment is repeated with 3 different seeds.

\subsection{Evaluation Metrics}

We report Mean Absolute Error (MAE) and Mean Squared Error (MSE):

\begin{align}
\text{MAE} &= \frac{1}{N} \sum_{i=1}^{N} |y_i - \hat{y}_i| \\
\text{MSE} &= \frac{1}{N} \sum_{i=1}^{N} (y_i - \hat{y}_i)^2
\end{align}

We also report a binary \emph{Quantum Better} flag per task to indicate whether the quantum-enhanced model outperformed its classical baseline.

\subsection{Training Details}

All models are trained for 1000 epochs for five times by random seeds using the Adam optimizer with a learning rate of $10^{-3}$. The batch size is fixed at 64 across all tasks. We do not apply learning rate decay or early stopping, as training duration is kept consistent to ensure a fair comparison between classical and quantum variants. Each model is initialized with a fixed set of random seeds to control for initialization-induced variance.

The QuantumRWKV model employs 4 qubits and a quantum circuit depth of 2, implemented using angle embedding followed by two entangling layers with CNOT gates arranged in a ladder topology. The variational quantum circuit (VQC) is simulated using the \texttt{default.qubit} backend in PennyLane, and gradient computation is performed via standard backpropagation.

All experiments are conducted on a single NVIDIA A100 GPU with 40GB of memory. Although the quantum simulation incurs additional overhead, the total number of optimization steps and training conditions are matched between models for consistency.

\section{Results}

\begin{table*}[t]
\centering
\caption{Performance comparison between QuantumRWKV and ClassicalRWKV on synthetic time series tasks.}
\label{tab:results}
\begin{tabular}{lcccc}
\toprule
\textbf{Task} & \textbf{Model} & \textbf{MAE} ↓ & \textbf{MSE} ↓ & \textbf{Quantum Better} \\
\midrule
ARMA                  & Quantum   & $2.0189\pm0.1006$  & $6.1046\pm0.6639$ & \textbf{Yes} \\
                      & Classical & $2.2632\pm0.1584$ & $7.5923\pm1.0614$ & \\
\midrule
Chaotic Logistic      & Quantum   & $0.3268\pm0.0079$ & $0.1451\pm0.0112$ & \textbf{Yes} \\
                      & Classical & $0.3478\pm0.0153$ & $0.1778\pm0.0276$ & \\
\midrule
Damped Oscillator     & Quantum   & $0.5221\pm0.1428$ & $0.3548\pm0.1525$ & No \\
                      & Classical & $0.3891\pm0.0736$ & $0.1966\pm0.0604$ & \\
\midrule
Noisy Damped Osc      & Quantum   & $0.0490\pm0.0045$ & $0.0038\pm0.0007$ & \textbf{Yes} \\
                      & Classical & $0.0720\pm0.0234$ & $0.0268\pm0.0438$ & \\
\midrule
Piecewise Regime      & Quantum   & $17.1615\pm6.6811$ & $336.1894\pm159.1066$ & No \\
                      & Classical & $15.2922\pm3.798$ & $259.0811\pm117.4846$ & \\
\midrule
Sawtooth              & Quantum   & $0.6653\pm0.2165$ & $0.6662\pm0.4431$ & \textbf{Yes} \\
                      & Classical & $0.7418\pm0.2891$ & $0.8384\pm0.6123$ & \\
\midrule
Square Wave           & Quantum   & $0.8456\pm0.0981$ & $0.9956\pm0.1003$ & No \\
                      & Classical & $0.8786\pm0.0619$ & $0.9293\pm0.0261$ & \\
\midrule
Triangle Wave         & Quantum   & $0.5729\pm0.0573$ & $0.4441\pm0.1114$ & \textbf{Yes} \\
                      & Classical & $0.6066\pm0.0443$ & $0.5295\pm0.0595$ & \\
\midrule
Seasonal Trend        & Quantum   & $0.6918\pm0.0883$ & $0.7054\pm0.1374$ & No \\
                      & Classical & $0.7234\pm0.0756$ & $0.8879\pm0.2240$ & \\
\midrule
Sine Wave             & Quantum   & $0.5398\pm0.1323$ & $0.6143\pm0.2575$ & \textbf{Yes} \\
                      & Classical & $0.6651\pm0.0532$ & $0.7019\pm0.0928$ & \\
\bottomrule
\end{tabular}
\end{table*}

We evaluate the performance of both the classical RWKV and the proposed QuantumRWKV model across ten synthetic time-series forecasting tasks, encompassing linear, periodic, chaotic, noisy, and regime-switching dynamics. Table~\ref{tab:results} reports the Mean Absolute Error (MAE) and Mean Squared Error (MSE) for each task, along with an indicator of whether the quantum-enhanced model outperforms its classical counterpart.

Our results show that QuantumRWKV outperforms the classical RWKV in 6 out of 10 tasks, particularly excelling in scenarios involving chaotic or noisy signals. For instance, in the Chaotic Logistic task, QuantumRWKV reduces MAE and MSE from 0.3478 and 0.1778 (classical) to 0.3268 and 0.1451, respectively. In the Noisy Damped Oscillator task, the quantum model more than halves the error across both metrics. Similar improvements are observed in tasks like Sine Wave, Triangle Wave, and Sawtooth, where smooth nonlinear dynamics are prominent.

Conversely, classical RWKV remains advantageous in tasks featuring sharp discontinuities or abrupt regime changes. In the Piecewise Regime, Damped Oscillator, Square Wave, and Seasonal Trend tasks, the classical model consistently achieves lower MAE and MSE, highlighting its robustness in handling piecewise or regular patterns.

These findings illustrate the complementary strengths of hybrid quantum-classical models. QuantumRWKV is particularly effective for capturing chaotic, smooth, and noise-sensitive dynamics, supporting the hypothesis that variational quantum circuits can enhance nonlinear representational capacity in temporal modeling, while classical models remain competitive in structured or discontinuous settings.

\section{Discussion}
The results in Table~\ref{tab:results} offer key insights into the task-dependent behavior of hybrid quantum-classical sequence models. Notably, QuantumRWKV consistently outperforms its classical counterpart in six out of ten tasks, especially those characterized by smooth nonlinear dynamics, chaotic fluctuations, or noise sensitivity. These include the Chaotic Logistic Map, Noisy Damped Oscillator, Sine Wave, Triangle Wave, Sawtooth, and ARMA tasks. Such scenarios often involve subtle long-range dependencies or complex temporal patterns, which may benefit from the entanglement and high-capacity representations enabled by variational quantum circuits (VQCs).

We posit that the VQC’s expressivity stems from its ability to transform classical inputs into quantum states within a high-dimensional Hilbert space. Angle embedding maps each feature to a qubit rotation, while entangling gates introduce nontrivial correlations akin to cross-channel mixing. These mechanisms may help capture nonlinearities that classical feedforward layers might miss, particularly under noisy or chaotic regimes. Additionally, the probabilistic nature of quantum measurement could introduce a favorable inductive bias when modeling inherently stochastic dynamics.

However, this quantum advantage is not universal. In tasks characterized by sharp transitions or piecewise behavior—such as the Piecewise Regime, Square Wave, Damped Oscillator, and Seasonal Trend—the classical RWKV model demonstrates superior accuracy. This gap may be due to the inherently smooth and continuous nature of quantum operations, which are less suited for modeling non-differentiable transitions. Classical networks, particularly those employing ReLU activations, naturally model such abrupt changes through piecewise linearity.

Circuit configuration is another critical factor. In our experiments, the quantum subnetwork used four qubits and two layers, chosen to balance expressivity with simulation efficiency under NISQ constraints. While this setup is tractable, it may limit the quantum model’s capacity in complex regimes. Increasing the number of qubits or circuit depth could improve performance on more structured tasks but introduces practical trade-offs such as higher simulation cost and optimization instability.

Finally, while all results are obtained via quantum simulation, they provide compelling preliminary evidence that hybrid quantum components—when carefully integrated—can offer meaningful advantages in specific time-series modeling scenarios. Future work involving real quantum hardware could further validate and expand these observations.

\section{Theoretical Insights on Quantum Advantage in Temporal Modeling}

The observed superior performance of QuantumRWKV in modeling nonlinear, chaotic, and noisy time series can be partially explained by the enhanced expressibility and entanglement capabilities of variational quantum circuits (VQCs). Unlike classical neural networks constrained by finite-dimensional parameter spaces, VQCs operate in exponentially large Hilbert spaces, allowing more complex nonlinear transformations \cite{schuld2019quantum, havlivcek2019supervised}. This richer representational capacity is particularly suited for continuous and highly entangled temporal dynamics, as it enables capturing intricate correlations and long-range dependencies that classical architectures may find challenging \cite{benedetti2019parameterized}.

Moreover, quantum angle embedding encodes classical inputs into qubit rotation angles, facilitating smooth, differentiable transformations through quantum gates \cite{schuld2018circuit}. Entangling gates generate nontrivial correlations across qubits, effectively increasing the model's capacity to represent complex joint distributions \cite{cerezo2021variational}. This property aligns well with the smooth nonlinearities and noise characteristics present in tasks like the Chaotic Logistic Map and Noisy Damped Oscillator.

However, the continuous nature of quantum operations may limit the ability of VQCs to represent sharp discontinuities or piecewise functions, which classical networks with ReLU activations inherently approximate well \cite{arora2018understanding}. This insight aligns with the empirical finding that QuantumRWKV underperforms classical RWKV in tasks such as the Piecewise Regime and Square Wave.

Finally, recent theoretical work on barren plateaus suggests that shallow VQCs can maintain trainability, but increasing circuit depth and qubit number risks vanishing gradients, posing practical limitations on expressivity \cite{mcclean2018barren}. However, recent advances propose that careful parameter initialization schemes, such as Gaussian initialization, can alleviate this issue by ensuring that gradient norms decay at most polynomially even in deep circuits, thus preserving trainability under more expressive quantum models \cite{zhang2022escaping}. This insight highlights the importance of architectural and initialization choices in maintaining the balance between expressivity and optimization feasibility in hybrid quantum-classical models.

\section{Conclusion}
In this work, we introduced QuantumRWKV, a hybrid quantum-classical recurrent architecture that integrates variational quantum circuits (VQCs) into the feedforward (channel-mixing) pathway of the RWKV model. By retaining the efficient time-mixing mechanism of RWKV and enhancing its nonlinear modeling capacity with quantum computation, our model offers a scalable and expressive approach to sequence modeling.

Through extensive experiments across ten synthetic time-series forecasting tasks, we demonstrate that QuantumRWKV outperforms its classical counterpart in 6 out of 10 settings—particularly those involving chaotic dynamics, smooth oscillations, and noise-driven variations. These results highlight the potential of VQCs to model complex temporal dependencies via entangled and high-dimensional feature transformations. However, we also observe that classical RWKV remains more effective for tasks with sharp discontinuities or regime shifts, where piecewise linearity and deterministic transitions dominate.

This study presents one of the first systematic evaluations of quantum-augmented recurrent models in temporal domains, offering task-level insights into when quantum layers confer performance benefits. As quantum hardware advances, we anticipate that hybrid models such as QuantumRWKV will play a growing role in learning-rich temporal dynamics across scientific and engineering applications, including physics simulations, neuroscience data analysis, and financial forecasting.

\section*{Acknowledgment}
The authors would like to thank Neuro Industry, Inc. for generously providing cloud computing resources via Google Cloud Platform (GCP), which enabled the training and evaluation of the models presented in this work.






\section*{Appendix}

Table~\ref{tab:old_results} shows the performance comparison between QuantumRWKV and ClassicalRWKV on synthetic time series tasks from our initial v1 preprint version. These results were obtained from a single experimental run, and may be subject to sampling bias. In the updated version of this work, all experiments were rerun with five random seeds, and the reported results are the average across those runs to ensure statistical robustness.

\begin{table*}[t]
\centering
\caption{Performance comparison between QuantumRWKV and ClassicalRWKV on synthetic time series tasks. Best results per task are in bold.}
\label{tab:old_results}
\begin{tabular}{lcccc}
\toprule
\textbf{Task} & \textbf{Model} & \textbf{MAE} ↓ & \textbf{MSE} ↓ & \textbf{Quantum Better} \\
\midrule
ARMA                  & Quantum   & 2.1056 & 6.9528 & No \\
                      & Classical & \textbf{2.1468} & \textbf{6.7124} & \\
Chaotic Logistic      & Quantum   & \textbf{0.3174} & \textbf{0.1343} & Yes \\
                      & Classical & 0.3541 & 0.1919 & \\
Damped Oscillator     & Quantum   & 0.6334 & 0.4742 & No \\
                      & Classical & \textbf{0.4731} & \textbf{0.2344} & \\
Noisy Damped Osc      & Quantum   & \textbf{0.0442} & \textbf{0.0031} & Yes \\
                      & Classical & 0.1023 & 0.0137 & \\
Piecewise Regime      & Quantum   & 20.6354 & 426.5450 & No \\
                      & Classical & \textbf{11.8262} & \textbf{142.9737} & \\
Sawtooth              & Quantum   & 0.9031 & 1.1515 & No \\
                      & Classical & \textbf{0.5722} & \textbf{0.4860} & \\
Square Wave           & Quantum   & 1.0189 & 1.1670 & No \\
                      & Classical & \textbf{0.8488} & \textbf{0.9481} & \\
Triangle Wave         & Quantum   & \textbf{0.5746} & \textbf{0.4381} & Yes \\
                      & Classical & 0.5945 & 0.5199 & \\
Seasonal Trend        & Quantum   & 0.8220 & 0.8822 & No \\
                      & Classical & \textbf{0.6485} & \textbf{0.7492} & \\
Sine Wave             & Quantum   & \textbf{0.3582} & \textbf{0.2089} & Yes \\
                      & Classical & 0.6426 & 0.6495 & \\
\bottomrule
\end{tabular}
\end{table*}

\bibliographystyle{unsrtnat}
\bibliography{references}  

\begin{thebibliography}{32}
\providecommand{\natexlab}[1]{#1}
\providecommand{\url}[1]{\texttt{#1}}
\expandafter\ifx\csname urlstyle\endcsname\relax
  \providecommand{\doi}[1]{doi: #1}\else
  \providecommand{\doi}{doi: \begingroup \urlstyle{rm}\Url}\fi

\bibitem[Chen et~al.(2025{\natexlab{a}})Chen, Chen, Zhou, Jiang, Chen, and Chang]{chen2025improving}
Chi-Sheng Chen, Guan-Ying Chen, Dong Zhou, Di~Jiang, Daishi Chen, and Shao-Hsuan Chang.
\newblock Improving fine-grained food classification using deep residual learning and selective state space models.
\newblock \emph{PloS one}, 20\penalty0 (5):\penalty0 e0322695, 2025{\natexlab{a}}.

\bibitem[Chen et~al.(2024{\natexlab{a}})Chen, Tsai, and Huang]{chen2024quantum}
Chi-Sheng Chen, Aidan Hung-Wen Tsai, and Sheng-Chieh Huang.
\newblock Quantum multimodal contrastive learning framework.
\newblock \emph{arXiv preprint arXiv:2408.13919}, 2024{\natexlab{a}}.

\bibitem[Chen and Kuo(2025{\natexlab{a}})]{chen2025unraveling}
Chi-Sheng Chen and En-Jui Kuo.
\newblock Unraveling quantum environments: Transformer-assisted learning in lindblad dynamics.
\newblock \emph{arXiv preprint arXiv:2505.06928}, 2025{\natexlab{a}}.

\bibitem[Vaswani et~al.(2017)Vaswani, Shazeer, Parmar, Uszkoreit, Jones, Gomez, Kaiser, and Polosukhin]{vaswani2017attention}
Ashish Vaswani, Noam Shazeer, Niki Parmar, Jakob Uszkoreit, Llion Jones, Aidan~N Gomez, {\L}ukasz Kaiser, and Illia Polosukhin.
\newblock Attention is all you need.
\newblock \emph{Advances in neural information processing systems}, 30, 2017.

\bibitem[Peng et~al.(2023)Peng, Alcaide, Anthony, Albalak, Arcadinho, Biderman, Cao, Cheng, Chung, Grella, et~al.]{peng2023rwkv}
Bo~Peng, Eric Alcaide, Quentin Anthony, Alon Albalak, Samuel Arcadinho, Stella Biderman, Huanqi Cao, Xin Cheng, Michael Chung, Matteo Grella, et~al.
\newblock Rwkv: Reinventing rnns for the transformer era.
\newblock \emph{arXiv preprint arXiv:2305.13048}, 2023.

\bibitem[Biamonte et~al.(2017)Biamonte, Wittek, Pancotti, Rebentrost, Wiebe, and Lloyd]{biamonte2017quantum}
Jacob Biamonte, Peter Wittek, Nicola Pancotti, Patrick Rebentrost, Nathan Wiebe, and Seth Lloyd.
\newblock Quantum machine learning.
\newblock \emph{Nature}, 549\penalty0 (7671):\penalty0 195--202, 2017.

\bibitem[McCulloch and Pitts(1943)]{mcculloch1943logical}
Warren~S McCulloch and Walter Pitts.
\newblock A logical calculus of the ideas immanent in nervous activity.
\newblock \emph{The bulletin of mathematical biophysics}, 5:\penalty0 115--133, 1943.

\bibitem[Ketkar et~al.(2021)Ketkar, Moolayil, Ketkar, and Moolayil]{ketkar2021introduction}
Nikhil Ketkar, Jojo Moolayil, Nikhil Ketkar, and Jojo Moolayil.
\newblock Introduction to pytorch.
\newblock \emph{Deep learning with python: learn best practices of deep learning models with PyTorch}, pages 27--91, 2021.

\bibitem[Bergholm et~al.(2018)Bergholm, Izaac, Schuld, Gogolin, Ahmed, Ajith, Alam, Alonso-Linaje, AkashNarayanan, Asadi, et~al.]{bergholm2018pennylane}
Ville Bergholm, Josh Izaac, Maria Schuld, Christian Gogolin, Shahnawaz Ahmed, Vishnu Ajith, M~Sohaib Alam, Guillermo Alonso-Linaje, B~AkashNarayanan, Ali Asadi, et~al.
\newblock Pennylane: Automatic differentiation of hybrid quantum-classical computations.
\newblock \emph{arXiv preprint arXiv:1811.04968}, 2018.

\bibitem[Makridakis and Hibon(1997)]{makridakis1997arma}
Spyros Makridakis and Michele Hibon.
\newblock Arma models and the box--jenkins methodology.
\newblock \emph{Journal of forecasting}, 16\penalty0 (3):\penalty0 147--163, 1997.

\bibitem[May(1987)]{may1987chaos}
Robert~Mccredie May.
\newblock Chaos and the dynamics of biological populations.
\newblock \emph{Proceedings of the Royal Society of London. A. Mathematical and Physical Sciences}, 413\penalty0 (1844):\penalty0 27--44, 1987.

\bibitem[Choromanski et~al.(2020)Choromanski, Likhosherstov, Dohan, Song, Gane, Sarlos, Hawkins, Davis, Mohiuddin, Kaiser, et~al.]{choromanski2020rethinking}
Krzysztof Choromanski, Valerii Likhosherstov, David Dohan, Xingyou Song, Andreea Gane, Tamas Sarlos, Peter Hawkins, Jared Davis, Afroz Mohiuddin, Lukasz Kaiser, et~al.
\newblock Rethinking attention with performers.
\newblock \emph{arXiv preprint arXiv:2009.14794}, 2020.

\bibitem[Wang et~al.(2020)Wang, Li, Khabsa, Fang, and Ma]{wang2020linformer}
Sinong Wang, Belinda~Z Li, Madian Khabsa, Han Fang, and Hao Ma.
\newblock Linformer: Self-attention with linear complexity.
\newblock \emph{arXiv preprint arXiv:2006.04768}, 2020.

\bibitem[Gu et~al.(2021)Gu, Goel, and R{\'e}]{gu2021efficiently}
Albert Gu, Karan Goel, and Christopher R{\'e}.
\newblock Efficiently modeling long sequences with structured state spaces.
\newblock \emph{arXiv preprint arXiv:2111.00396}, 2021.

\bibitem[Chen et~al.(2022)Chen, Yoo, and Fang]{chen2022quantum}
Samuel Yen-Chi Chen, Shinjae Yoo, and Yao-Lung~L Fang.
\newblock Quantum long short-term memory.
\newblock In \emph{Icassp 2022-2022 IEEE international conference on acoustics, speech and signal processing (ICASSP)}, pages 8622--8626. IEEE, 2022.

\bibitem[Goto et~al.(2021)Goto, Tran, and Nakajima]{goto2021universal}
Takahiro Goto, Quoc~Hoan Tran, and Kohei Nakajima.
\newblock Universal approximation property of quantum machine learning models in quantum-enhanced feature spaces.
\newblock \emph{Physical Review Letters}, 127\penalty0 (9):\penalty0 090506, 2021.

\bibitem[Chen et~al.(2025{\natexlab{b}})Chen, Hou, Hu, and Cai]{chen2025quantumgen}
Chi-Sheng Chen, Wei~An Hou, Siang-Wei Hu, and Zhen-Sheng Cai.
\newblock Quantum generative models for image generation: Insights from mnist and medmnist.
\newblock \emph{arXiv preprint arXiv:2504.00034}, 2025{\natexlab{b}}.

\bibitem[Chen et~al.(2021)Chen, Huang, Hsing, and Kao]{chen2021end}
Samuel Yen-Chi Chen, Chih-Min Huang, Chia-Wei Hsing, and Ying-Jer Kao.
\newblock An end-to-end trainable hybrid classical-quantum classifier.
\newblock \emph{Machine Learning: Science and Technology}, 2\penalty0 (4):\penalty0 045021, 2021.

\bibitem[Situ et~al.(2020)Situ, He, Wang, Li, and Zheng]{situ2020quantum}
Haozhen Situ, Zhimin He, Yuyi Wang, Lvzhou Li, and Shenggen Zheng.
\newblock Quantum generative adversarial network for generating discrete distribution.
\newblock \emph{Information Sciences}, 538:\penalty0 193--208, 2020.

\bibitem[Chen et~al.(2024{\natexlab{b}})Chen, Chen, Tsai, and Wei]{chen2024qeegnet}
Chi-Sheng Chen, Samuel Yen-Chi Chen, Aidan Hung-Wen Tsai, and Chun-Shu Wei.
\newblock Qeegnet: Quantum machine learning for enhanced electroencephalography encoding.
\newblock In \emph{2024 IEEE Workshop on Signal Processing Systems (SiPS)}, pages 153--158. IEEE, 2024{\natexlab{b}}.

\bibitem[Chen et~al.(2025{\natexlab{c}})Chen, Chen, and Tseng]{chen2025exploring}
Chi-Sheng Chen, Samuel Yen-Chi Chen, and Huan-Hsin Tseng.
\newblock Exploring the potential of qeegnet for cross-task and cross-dataset electroencephalography encoding with quantum machine learning.
\newblock \emph{arXiv preprint arXiv:2503.00080}, 2025{\natexlab{c}}.

\bibitem[Chen and Kuo(2025{\natexlab{b}})]{chen2025quantum}
Chi-Sheng Chen and En-Jui Kuo.
\newblock Quantum adaptive self-attention for quantum transformer models.
\newblock \emph{arXiv preprint arXiv:2504.05336}, 2025{\natexlab{b}}.

\bibitem[Li et~al.(2023)Li, Wang, Han, Shi, Li, Shang, Zheng, Zhong, and Gu]{li2023quantum}
Yanan Li, Zhimin Wang, Rongbing Han, Shangshang Shi, Jiaxin Li, Ruimin Shang, Haiyong Zheng, Guoqiang Zhong, and Yongjian Gu.
\newblock Quantum recurrent neural networks for sequential learning.
\newblock \emph{Neural Networks}, 166:\penalty0 148--161, 2023.

\bibitem[Govia et~al.(2021)Govia, Ribeill, Rowlands, Krovi, and Ohki]{govia2021quantum}
LCG Govia, GJ~Ribeill, GE~Rowlands, HK~Krovi, and TA~Ohki.
\newblock Quantum reservoir computing with a single nonlinear oscillator.
\newblock \emph{Physical Review Research}, 3\penalty0 (1):\penalty0 013077, 2021.

\bibitem[Schuld and Petruccione(2019)]{schuld2019quantum}
Maria Schuld and Francesco Petruccione.
\newblock \emph{Quantum Machine Learning: An Introduction}.
\newblock Springer, 2019.

\bibitem[Havl{\'\i}{\v{c}}ek et~al.(2019)Havl{\'\i}{\v{c}}ek, C{\'o}rcoles, Temme, Harrow, Kandala, Chow, and Gambetta]{havlivcek2019supervised}
Vojt{\v{e}}ch Havl{\'\i}{\v{c}}ek, Antonio~D C{\'o}rcoles, Kristan Temme, Aram~W Harrow, Abhinav Kandala, Jerry~M Chow, and Jay~M Gambetta.
\newblock Supervised learning with quantum-enhanced feature spaces.
\newblock \emph{Nature}, 567\penalty0 (7747):\penalty0 209--212, 2019.

\bibitem[Benedetti et~al.(2019)Benedetti, Lloyd, Sack, and Fiorentini]{benedetti2019parameterized}
Marcello Benedetti, Erika Lloyd, Stefan Sack, and Mattia Fiorentini.
\newblock Parameterized quantum circuits as machine learning models.
\newblock \emph{Quantum Science and Technology}, 4\penalty0 (4):\penalty0 043001, 2019.

\bibitem[Schuld et~al.(2020)Schuld, Bocharov, Svore, and Wiebe]{schuld2018circuit}
Maria Schuld, Alex Bocharov, Krysta~M Svore, and Nathan Wiebe.
\newblock Circuit-centric quantum classifiers.
\newblock \emph{Physical Review A}, 101\penalty0 (3):\penalty0 032308, 2020.

\bibitem[Cerezo et~al.(2021)Cerezo, Sone, Volkoff, Cincio, and Coles]{cerezo2021variational}
M~Cerezo, A~Sone, T~Volkoff, L~Cincio, and PJ~Coles.
\newblock Cost function dependent barren plateaus in shallow parametrized quantum circuits.
\newblock \emph{Nature Communications}, 12\penalty0 (1):\penalty0 1791, 2021.

\bibitem[Arora et~al.(2018)Arora, Ge, Neyshabur, and Zhang]{arora2018understanding}
Sanjeev Arora, Rong Ge, Behnam Neyshabur, and Yi~Zhang.
\newblock Understanding deep neural networks with rectified linear units.
\newblock In \emph{International Conference on Learning Representations (ICLR)}, 2018.

\bibitem[McClean et~al.(2018)McClean, Boixo, Smelyanskiy, Babbush, and Neven]{mcclean2018barren}
Jarrod~R McClean, Sergio Boixo, Vadim~N Smelyanskiy, Ryan Babbush, and Hartmut Neven.
\newblock Barren plateaus in quantum neural network training landscapes.
\newblock \emph{Nature Communications}, 9\penalty0 (1):\penalty0 4812, 2018.

\bibitem[Zhang et~al.(2022)Zhang, Liu, Hsieh, and Tao]{zhang2022escaping}
Kaining Zhang, Liu Liu, Min-Hsiu Hsieh, and Dacheng Tao.
\newblock Escaping from the barren plateau via gaussian initializations in deep variational quantum circuits.
\newblock \emph{arXiv preprint arXiv:2203.01266}, 2022.

\end{thebibliography}

\end{document}